\documentclass[reqno,11pt]{amsart}
\usepackage{psfig, amsmath, amsfonts, amssymb, amsthm, amscd,enumerate}

\DeclareMathSymbol{\leqslant}{\mathalpha}{AMSa}{"36} % nicer `smaller or equal' \DeclareMathSymbol{\geqslant}{\mathalpha}{AMSa}{"3E} % nicer `larger or equal'
\DeclareMathSymbol{\eset}{\mathalpha}{AMSb}{"3F}     % nicer `emptyset'
\renewcommand{\leq}{\;\leqslant\;}                   % redef. of < or =
\renewcommand{\geq}{\;\geqslant\;}                   % redef. of > or =
\makeatletter
\def\captionfont@{\footnotesize} \def\captionheadfont@{\scshape}

\long\def\@makecaption#1#2{%
  \vspace{2mm}
  \setbox\@tempboxa\vbox{\color@setgroup
    \advance\hsize-6pc\noindent
    \captionfont@\captionheadfont@#1\@xp\@ifnotempty\@xp
        {\@cdr#2\@nil}{.\captionfont@\upshape\enspace#2}%
    \unskip\kern-6pc\par
    \global\setbox\@ne\lastbox\color@endgroup}%
  \ifhbox\@ne % the normal case
    \setbox\@ne\hbox{\unhbox\@ne\unskip\unskip\unpenalty\unkern}%
  \fi
  \ifdim\wd\@tempboxa=\z@ % this means caption will fit on one line
    \setbox\@ne\hbox to\columnwidth{\hss\kern-6pc\box\@ne\hss}%
  \else % tempboxa contained more than one line
    \setbox\@ne\vbox{\unvbox\@tempboxa\parskip\z@skip
        \noindent\unhbox\@ne\advance\hsize-6pc\par}%
\fi
  \ifnum\@tempcnta<64 % if the float IS a figure...
    \addvspace\abovecaptionskip
    \moveright 3pc\box\@ne
  \else % if the float IS NOT a figure...
    \moveright 3pc\box\@ne
    \nobreak
    \vskip\belowcaptionskip
  \fi
\relax
}
\makeatother
\def\writefig#1 #2 #3 {\rlap{\kern #1 truecm
\raise #2 truecm \hbox{#3}}}

\psfigurepath{.:./pictures} %%%%%%%%%%%%%%%%%%%%%%%%%%%%%%%%%%%%%%%%%%%%%%%%%%%%%%%%%%%%%%%%%%%%%%%%%%%%%%
\newtheorem{lem}{Lemma}[section] \newtheorem{pro}{Proposition}[section]
\newtheorem{thm}{Theorem}[section]

\newtheorem{rem}{Remark}[section] \newtheorem{definition}{Definition}[section]
 \newcommand{\cB}{\ensuremath{\mathcal B}} \newcommand{\cC}{\ensuremath{\mathcal C}} \newcommand{\cD}{\ensuremath{\mathcal D}} \newcommand{\cE}{\ensuremath{\mathcal E}} \newcommand{\cF}{\ensuremath{\mathcal F}}  \newcommand{\cH}{\ensuremath{\mathcal H}}    \newcommand{\cL}{\ensuremath{\mathcal L}} \newcommand{\cM}{\ensuremath{\mathcal M}}  \newcommand{\cO}{\ensuremath{\mathcal O}}  \newcommand{\cQ}{\ensuremath{\mathcal Q}}     \newcommand{\cV}{\ensuremath{\mathcal V}} \newcommand{\cW}{\ensuremath{\mathcal W}}   

 \newcommand{\bbB}{{\ensuremath{\mathbb B}} }   \newcommand{\bbE}{{\ensuremath{\mathbb E}} }       \newcommand{\bbL}{{\ensuremath{\mathbb L}} }      \newcommand{\bbR}{{\ensuremath{\mathbb R}} } \newcommand{\bbS}{{\ensuremath{\mathbb S}} }    \newcommand{\bbW}{{\ensuremath{\mathbb W}} }   \newcommand{\bbZ}{{\ensuremath{\mathbb Z}} }

\newcommand{\ga}{\alpha}
\newcommand{\gb}{\beta}
\newcommand{\gga}{\gamma}            % \gg already exists...
\newcommand{\gd}{\delta}
\newcommand{\gep}{\varepsilon}       % \ge already exists...

\newcommand{\gr}{\rho}
\newcommand{\gz}{\zeta}

\newcommand{\gk}{\kappa}
\newcommand{\go}{\omega}

\newcommand{\gl}{\lambda}
\newcommand{\gL}{\Lambda}
\newcommand{\gs}{\sigma}

\newcommand{\gt}{\vartheta}

\let\a=\alpha \let\b=\beta   \let\d=\delta  
 \let\g=\gamma \let\h=\eta      \let\l=\lambda
      \let\o=\omega      
  \let\s=\sigma \let\t=\tau   
  
     \let\L=\Lambda 
\let\O=\Omega

\newcommand{\Is}{\mu}              % Ising measure
\newcommand{\Tor}[1]{\bbW_{#1}}    % Torus of size #1
\newcommand{\uTor}{\widehat \bbW^d}         % Unit torus
\def\1{\ifmmode {1\hskip -3pt \rm{I}} \else {\hbox {$1\hskip -3pt \rm{I}$}}\fi}
\newcommand{\BV}{{ {\rm BV}(\uTor,\{\pm1\}) }}         % BV

\newcommand{\sTor}[1]{\widehat{\bbW}_{#1}}   % Lattice torus of step 2^{-#1}
\newcommand{\sBox}[1]{\widehat{\bbB}_{#1}} %continuous box of size 2^{-#1} 
\newcommand{\dBox}[1]{{\bbB}_{#1}} %continuous box of size 2^{-#1} 
 
\newcommand{\st}{\tau}

 \let\mmax=\vee

\begin{document}
\title{Some New Results on the Kinetic Ising Model in a Pure Phase}

\author[T.~Bodineau]{T.~Bodineau}
\address{D{\'e}partement de math{\'e}matiques, Universit{\'e} Paris 7, case 7012, 2 place Jussieu, Paris 75251, France} \email{bodineau@gauss.math.jussieu.fr}
\author[F. Martinelli]{Fabio Martinelli}
\address{Dip. Matematica, Universita' di Roma Tre, L.go
S. Murialdo 1, 00146 Roma, Italy}
\email{martin\@@mat.uniroma3.it}

\thanks{During the first part of this work we have benefit from instructive conversations with D. Ioffe whom we warmly thank.
The second part
of this work was done while both authors were visiting the Institut H. Poincar{\'e} during the special semester devoted to ``Hydrodynamic Limits''. We would like to warmly thank the organizers, S. Olla and F. Golse, for their hospitality there and for the very stimulating scientific atmosphere.
T. B. also acknowledges Universities of Roma II and Roma III  for their invitation when this work started. Finally, we would like to thank Y. Velenik and N. Yoshida for useful comments.}

\begin{abstract}
We consider a general class of Glauber dynamics reversible with respect to the standard Ising model in $\bbZ^d$ with zero external field and inverse temperature $\gb$ strictly larger than the critical value $\gb_c$ in dimension 2 or the so called ``slab threshold'' $\hat \b_c$ in dimension $d \geq 3$.  We first prove that the inverse spectral gap in a large cube of side $N$ with plus boundary conditions is, apart from logarithmic corrections, larger than $N$ in $d=2$ while the logarithmic Sobolev constant is instead larger than $N^2$ in any dimension. Such a result substantially improves over all the previous existing bounds and agrees with a similar computations obtained in the framework of a one dimensional toy model based on mean curvature motion. The proof, based on a suggestion made by H.T. Yau some years ago, explicitly constructs a subtle test function which forces a large droplet of the minus phase inside the plus phase. The relevant bounds for general $d\ge 2$ are then obtained via a careful use of the recent $\bbL^1$--approach to the Wulff construction. Finally we prove that in $d=2$ the probability that two independent initial configurations, distributed according to the infinite volume plus phase and evolving under any coupling, agree at the origin at time $t$ is bounded from below by a stretched exponential $\exp(-\sqrt{t})$, again apart from logarithmic corrections. Such a result should be considered as a first step toward a rigorous proof that, as conjectured by Fisher and Huse some years ago, the equilibrium time auto-correlation of the spin at the origin decays as a stretched exponential in $d=2$.  \vskip.8cm



\end{abstract}

\maketitle

\section{Introduction}
In a finite domain, the reversible Glauber dynamics associated to the Ising model relaxes exponentially fast to its equilibrium measure. Nevertheless, this simple statement hides a wide range of  behaviors
depending on the temperature, the domain and the boundary conditions.

In the uniqueness regime (when the temperature is large enough), the speed of relaxation is uniform with respect to the domains and the boundary conditions.  We refer to Martinelli \cite{M1} for a complete account of this theory.  The occurrence of phase transition drastically modifies the behavior of the dynamics and new physical features slow down the relaxation; among those, the nucleation and the interface motions.  Metastability is characteristic of these slow phenomena since the system is trapped for a very long period of time in a local equilibrium.  In this case, the relaxation mechanism is so slow that the time of nucleation can be expressed in terms of equilibrium quantities. In particular, it was proven by Martinelli (see eg. \cite{M1} and references therin) that for free boundary conditions the asymptotic of the spectral gap with respect to the size of the domains is related to the surface tension and the main mechanism driving the system to equilibrium is nucleation of one phase inside the other. A complete picture of the nucleation process in $\bbZ^2$ in the framework of metastability was obtained by Schonmann and Shlosman in \cite{SS2}.\\

In this paper, we are interested in a different regime in which the relaxation to equilibrium is driven by the slow motion of the interfaces. This is the case of the Ising model in a large box with plus boundary conditions. When a droplet of the minus phase is surrounded by the plus phase, it tends to shrink according to its curvature under the action of the non-conservative dynamics on the spins close to the interface. This subtle phenomenon has been studied rigorously only in rare instances: by Spohn \cite{Spohn} in the case of Ising model at zero temperature (see also Rezakhanlou, Spohn \cite{RS}), by Chayes, Schonmann, Swindle \cite{CSS} for a variant of this model and by De Masi, Orlandi, Presutti, Triolo \cite{DOPT1, DOPT2} for the Kac-Ising model.  Notice also that the motion by mean curvature plays a key role in the coarsening phenomenon, as it has been shown recently by Fontes, Schonmann, Sidoravicius \cite{FSS}.  For positive temperatures, a mathematical derivation of similar results seems to be more challenging.\\

A way to capture some insights into the slow relaxation driven by interface motion is to estimate spectral quantities related to the generator of the Glauber dynamics.  We prove that for any dimension $d \geq 2$, in the phase transition regime and with plus boundary conditions, the logarithmic-Sobolev constant for a domain of linear size $N$ diverge at least like $N^2$ (up to some logarithmic corrections). This can be considered as a first characterization of the slow down of the dynamics and is in agreement with the heuristics predicted by the motion by mean curvature. In the same setting but $d=2$, we prove that the inverse of the spectral gap grows at least like $N$ (up to logarithmic corrections). In dimension $d \geq 3$ our argument fails to produce a result on the divergence of the spectral gap.\\

Let us stress that we have not been able to derive matching upper bounds; the best existing bounds have been proved only in $d=2$ and are of the form $\exp\bigl(\sqrt{N}(\log N)^2\bigr)$ (see \cite{YW}). However, an exact computation for a toy model based on mean curvature motion seems to confirm that the polynomial asymptotics we obtain are correct (see Section \ref{sec: toy model}).  The proof boils down to bound the variational formula for the Poincar{\'e} and the Log-Sobolev inequalities by choosing an appropriate test function. This reduces the problem to a computation under the equilibrium Gibbs measure.  The main difficulty is to recover polynomial bounds by using only the exponential estimates provided by the equilibrium theory of phase segregation (see \cite{BIV} and references therein).  This is achieved by the choice of a subtle test function which was suggested some years ago by H.T.~Yau.\\

The second part of the paper (section \ref{sec: Glauber dynamics}) applies the result on the lower bound on the inverse of the spectral gap
to investigate the relaxation in the infinite domain $\bbZ^2$. Thanks to an heuristic argument based on the motion by mean curvature, Fisher and Huse \cite{HF} conjectured that the equilibrium time auto-correlation of the spin at the origin decays as a stretched exponential  $\exp(-\sqrt{t})$ in $d=2$. We provide a first step towards a rigorous proof by showing that a dynamical quantity strictly related to the auto-correlation cannot relax faster than $\exp(-\sqrt{t})$.

\section{The model and the Main Results}

In this section we define the model and fix some useful the notation, recall some basic facts about the Ising model below the critical point and finally state our two main results.

\subsection{The standard Ising model}
Let $\L$ be a generic finite subset of $\bbZ^d$, with $d \geq 2$.  Each site $i$ in $\L$ indexes a spin $\gs_i$ which takes values $\pm 1$.  The spin configurations $\{ \gs_i \}_{i \in \L}$ have a statistical weight determined by the Hamiltonian \begin{equation*} H^{\bar \gs} ( \gs ) =
- \frac{1}{2} \sum_{i,j \in \L \atop |i-j|=1} \gs_i \gs_j
- \sum_{i \in \L, j  \in \L^c \atop |i-j|=1}\gs_i \bar \gs_j \, , \end{equation*} where $\bar \gs = \{ \bar \gs_i \}_{i \in \L^c}$ are boundary conditions
outside $\L$.

The Gibbs measure associated to the spin system with boundary conditions $\bar \gs$ is \begin{equation*} \forall \gs = \{ \gs_x\}_{x \in \L}, \qquad \Is^{\bar \gs}_\gL ( \gs ) = \frac{1}{Z_{\gb,\L}^{\bar \gs}} \exp \left( - \gb H^{\bar \gs} ( \gs ) \right), \end{equation*} where $\gb$ is the inverse of the temperature ($\gb = \frac{1}{T}$) and
$Z_{\gb,\L}^{\bar \gs}$ is the partition function.
If the boundary conditions are uniformly equal to 1 (resp. $-1$), the Gibbs measure will be denoted by $\Is_\L^+$ (resp. $\Is_\L^-$).

The phase transition regime occurs at low
temperature and is characterized by spontaneous magnetization
in the thermodynamic limit.
There is a critical value $\gb_c$ such that
\begin{equation}
\label{magnetization}
\forall \gb > \gb_c, \qquad
\lim_{\L\to \bbZ^d} \Is_\L^+ (\gs_0) = -
\lim_{\L\to \bbZ^d} \Is_\L^- (\gs_0) = m^* >0 \, . \end{equation} Furthermore, in the thermodynamic limit the measures $\Is_\L^+$ and $\Is_\L^-$ converge (weakly) to two distinct Gibbs measures $\mu^+$ and $\mu^-$ which are measures on the space $\{ \pm 1\}^{\bbZ^d}$.  Each of these measures represents a pure state.  In dimension $d \geq 3$, we also denote by $\hat \gb_c \geq \gb_c$ the ``slab critical point'' (see \cite{ACCFR} and \cite{pisztora}) which is conjectured to coincide with $\gb_c$. For convenience we set $\hat \gb_c = \gb_c$ in dimension 2. Our proofs rely on results of equilibrium phase coexistence for the Ising model which are restricted to values $\gb >\hat \gb_c$ (for technical reasons).\\

The next step is to quantify the coexistence of the two pure states defined above. Due to the lattice structure, the surface tension is an-isotropic. Let $\L= \{-N, \dots ,N\}^d$, let $\vec{n}$ be a vector in $\bbS^{d-1}$ such that $\vec{n} \cdot \vec{e}_1 >0$ and let $\bar \gs$ be the following mixed boundary conditions \begin{equation*} \forall i \in \L^c, \qquad \bar \gs_i  =
\begin{cases}
+1, & \qquad \text{if} \quad  \vec{n} \cdot i \geq 0, \\
-1, & \qquad \text{if} \quad  \vec{n} \cdot i <  0 . \end{cases} \end{equation*} The partition function with mixed boundary conditions is denoted by $Z_{\gb, N}^{\pm} (\vec{n})$ and the one with boundary conditions uniformly equal to $+1$ by $Z_{\gb, N}^+$.

\begin{definition}
The surface tension in the direction $\vec{n} \in \bbS^{d-1}$, with
$\vec{n} \cdot \vec{e}_1 >0$, is defined by
\begin{equation}
\label{surface tension}
\tau(\vec{n}) =  \lim_{N \to \infty} \, - \frac{(\vec{n},\vec{e}_1)}{N^{d-1}}
\,
\log \frac{Z_{\gb, N}^{\pm} (\vec{n})}{Z_{\gb, N}^+} \, . \end{equation} \end{definition}

We refer to Messager, Miracle-Sol{\'e} and Ruiz \cite{MMR} for a derivation of the thermodynamic limit \eqref{surface tension}. Associated in a natural way to the surface tension is the Wulff shape which describes the optimal shape of a droplet of the minus phase immersed in the plus phase. \begin{definition} The Wulff shape is the convex set in $\bbR^d$ given by \begin{equation} \label{Wulff shape} {\bf W} = \bigcap_{\vec{n} \in \bbS^{d-1}} \left\{ x \in \bbR^d; \qquad  x \cdot \vec{n} \leq \tau(\vec{n})
\right\} \, .
\end{equation}
\end{definition}
The Wulff shape with volume $1$  is denoted by
$\uTor$.
 Finally in what
follows we will choose for simplicity the finite set $\L$ as the domain
$\Tor{N} = N \uTor \cap \bbZ^d$, instead of a cube of side $N$.
The corresponding Gibbs measure on $\Tor{N}$ with $+$ boundary conditions
will be denoted by $\mu^+_N$. \\

\subsection{ The Glauber dynamics}

The stochastic dynamics we want to study is defined by
the Markov generator given by
$$
   (\cL_N^+ f)(\s) = \sum_{x\in \Tor{N}} c^+_x(\s)\nabla_x f(\s) $$
where the values of $\s$ outside $\Tor{N}$ are kept fixed identical to $+1$ and $\nabla_x f(\s) =\left[ f(\s^{x}) - f(\s) \right]$. On the flip rates $c_x(\s)$ we assume \begin{enumerate}[(i)] \item $k^{-1} \le c^+_x(\s) \le k$ for some $k$ and any $x,\s$ \item reversibility w.r.t. the Gibbs measure $\mu^+_N$ \item finite range
\end{enumerate}
\begin{rem}
It is possible to check (see e.g. \cite{Li} or \cite{M1}) that it is possible to extend the above definition of the generator $\cL_N^+$ directly to the whole lattice $\bbZ^d$ and get a well defined Markov process on $\O :=\{0,1\}^{\bbZ^d}$. We will refer to the latter as the infinite volume Glauber dynamics. \end{rem} The Dirichlet form associated to $\cL_N^+$ takes the form $$
\cE_N^+(f,f) = \sum_{x\in \Tor{N}} \mu^+_N\bigl(\, c_x(\s)
|\nabla_x f|^2\,\bigr)
$$
and, thanks to assumption (i) on the flip rates it is uniformly bounded from above and from below by $$ \mu^+_N\bigl(\, \sum_{x\in \Tor{N}} |\nabla_x f|^2\,\bigr) :=
\mu^+_N\bigl(\,|\nabla f|^2\,\bigr)
$$
Two key quantities measure the time scale on which relaxation to
equilibrium occurs. The first one, denoted by $S_N$, is the inverse of the spectral gap of the generator, while the other one the logarithmic Sobolev constant $L_N$. They are both characterized by a variational principle in that they are the optimal constants in the Poincar{\'e} inequality $$
\mu_N^+(f,f) \le c \,\cE_N^+(f,f), \qquad \forall \, f\in L^2(d\mu_N^+) $$ and in the logarithmic Sobolev inequality $$  \mu_N^+(f^2\log f^2) \le c \,\cE_N^+(f,f) \, , \qquad
\forall \, f \in L^2(d\mu_N^+) \ \ \text{ with } \mu_N^+(f^2)=1 $$
respectively. As it is well known the quantity $S_N$ measures the relaxation time in an $L^2(d\mu_N^+)$ sense while $L_N$ measures the relaxation time in an $L^\infty$ sense (worst case for the initial condition). More precisely, if $P^{(+,N)}_t$ denotes the Markov semigroup generated by $\cL_N^+$ and $f$ is an arbitrary function with $\mu_N^+(f)=0$ then \begin{align*}
  \mu_N^+\bigl(\,[P^{(+,N)}_tf]^2\,\bigr) &\le \mu_N^+(f^2) \exp(-\frac{t}{S_N}) \, . \end{align*}

In many cases e.g. at high temperature the two quantities
are of the same order but it may very well happen that they are quite different. We will argue later on that the Ising model below the critical temperature is actually one of these cases.

\subsection{Main Results}
\label{subsec: Main Results}
We are finally in a position to state our main results.

\begin{thm}
\label{pro spectral gap}
Assume $d=2$ and $\b > \b_c$. There exists a constant $\gk$ depending on $\gb$ such that \begin{equation} \lim_{N \to \infty} \frac{(\log N)^\gk}{N} S_N = +\infty
\end{equation}
\end{thm}
\begin{rem}
As we already pointed out in the introduction, in dimension greater than two our choice of the test function to be inserted in the Poincar{\'e} inequality does not provide any non trivial information. \end{rem}
The next
result concerns the large $N$ behavior of the logarithmic Sobolev constant. \begin{thm} \label{pro log sob} Assume $d\geq 2$ and $\b > \hat \b_c$. There exists a constant $\gk$ depending on $\gb$ and $d$ such that \begin{equation} \lim_{N \to \infty} \frac{(\log N)^\gk}{N^2} L_N = +\infty
\end{equation}
\end{thm}

\medskip

Finally we investigate in $d=2$ the relaxation in the plus phase for the infinite volume dynamics. For this purpose, let us consider an arbitrary coupling of the Glauber dynamics in the infinite volume $\bbZ^2$. The two processes at time $t$ are denoted by $(\gs^\eta (t), {\tilde \gs}^\go(t))$, where $(\eta,\go)$ are the initial spin configurations. The joint expectation of the process is denoted by $\hat \bbE$.  The initial conditions will in general be chosen w.r.t. the product measure $d {\hat \mu}^+ (\eta, \go) = d\mu^+ (\eta) d\mu^+ (\go)$, where $\mu^+$ is the Gibbs measure in the $+$ pure phase.

\begin{thm}
\label{stretch decay}
There exist positive constants $C_1, C_2$ and $\gk$ independent of the choice of the coupling such that \begin{eqnarray}
\label{eq: stretch decay}
\forall t >0, \qquad
\int \, d {\hat \mu}^+ (\eta, \go) \,
{\hat \bbE} \big( \gs_0^\eta (t) \neq {\tilde \gs}_0^\go (t)
\big)
\geq C_1 \exp\big( - C_2 \sqrt{t}\, (\log t)^\gk \big)
\, .
\end{eqnarray}
\end{thm}

\begin{rem}
Although we believe that the quantity considered in the theorem is a good measure of the time auto--correlation in the plus phase of the spin at the origin, the latter is unfortunately only bounded from {\sl above} by the LHS of \eqref{eq: stretch decay}. We have in fact \begin{eqnarray*} && \mu^+ \left( \big( P_t (\gs_0) - m^* \big)^2 \right) = \mu^+
\left( \big( P_t (\gs_0) - \int \, d\mu^+ {\tilde P_t}(\gs_0)
\big)^2 \right) \\
&& \qquad = \int \, d\mu^+ (\eta) \left(
\left( \int \, d\mu^+ (\go) \, {\hat \bbE}
\big( \gs_0^\eta (t) -  {\tilde \gs}_0^\go(t)  \big) \right)^2
\right) \leq 4
\int \, d {\hat \mu}^+ (\eta, \go) {\hat \bbE}
\big( \gs_0^\eta (t) \neq  {\tilde \gs}_0^\go(t) \big) \, . \end{eqnarray*} \end{rem} \begin{rem} A related result at $\b=+\infty$ was proved recently in \cite{FSS}
for the zero temperature dynamics (see theorem 1.2 there).
\end{rem}

\section{ Large Deviations}
\label{sec: Large Deviations}

\noindent
In this section we recall some results on the large deviations for the Gibbs measure $\mu_N^+$ when $\b > \hat \b_c$. Our proofs rely on a weak description of phase segregation in terms of $\bbL^1$--norm.
In dimension 2, more precise results can be found in Ioffe, Schonmann
\cite{IS} and in Pfister, Velenik \cite{PfisterVelenik97} (for
$\bbL^1$-concentration statements).
The reader is referred to \cite{BIV} for a survey on phase coexistence and a complete list of references.\\

\noindent
We consider our microscopic Ising model embedded in $\uTor$.  Let $\sTor{N} = \frac{1}{N} \bbZ^d \cap \uTor$ and let $K$ be a mesoscopic scale (eventually depending on $N$). The domain $\uTor$ is partitioned into boxes $\sBox{N,K}$, each of them containing $K^d$ sites of $\sTor{N}$: \begin{equation*} j \in \bbZ^d, \ x_j = j \frac{K}{N} \in \sTor{N}, \qquad
\sBox{N,K}(x_j) = x_j + \left]-\frac{K}{2 N}, \frac{K}{2 N} \right]^d \, . \end{equation*} Let $\dBox{K}(Nx_j)$ be the microscopic counterpart of $\sBox{N,K}(x_j)$, i.e. the sites of $\sTor{N}$ in $\sBox{N,K}(x_j)$. These boxes are centered on the sites of
$\sTor{N,K} = \frac{K}{N} \bbZ^d \cap \uTor$.
As the domain is not regular some boxes may not fit inside $\sTor{N}$, therefore at the boundary we consider a  relaxed notion of boxes. \\ Finally, the local magnetization is defined as a piece-wise constant function on the partition $\{ \sBox{N,K}(x_j) \}$:
\begin{equation}
\label{eq local magnetization}
\forall y \in \sBox{N,K}(x_j), \qquad
\cM_{N,K}(y) = \frac{1}{|\dBox{K}|} \sum_{i \in \dBox{K}(Nx_j)} \gs_i \, . \end{equation} The local order parameter $\cM_{N,K}(y)$ characterizes the local equilibrium of the mesoscopic box containing $y$. The key result concerning the local order parameters is a trivial consequence of the results obtained by Pisztora \cite{pisztora} and it is based on the following coarse grained description.  To each box $\sBox{N,K}(x_j)$ we associate a mesoscopic phase label $u_{N,K}(x_j)$ taking values in $\{-1,0,1\}$ \begin{eqnarray*} \label{labels}
u_{N,K}(x_j) =
1_{ \{ | \cM_{N,K}(x_j) - m^* | \leq \frac{1}{4} m^* \} }
-
1_{ \{ | \cM_{N,K}(x_j) + m^* | \leq \frac{1}{4} m^*  \} }  \, . \end{eqnarray*} The distribution of the variables $\{u_{N,K}(x_j)\}$ under $\mu_N^+$ is dominated by Bernoulli Percolation.\\

\begin{thm}[\cite{pisztora}] For any $\b > \hat \b_c$ there exists
$c_\gb >0$ and $\gga \in \,]0,1[$ such that the following holds uniformly in $N$: \begin{equation} \label{eq domination} \forall \{x_1, \dots, x_\ell \} \in \sTor{N,K}, \quad \mu^+_N \left(
u_{N,K}(x_1)  = 0, \dots , u_{N,K}(x_\ell) = 0
\right)
\leq \big( \gr_K \big)^\ell \, ,
\end{equation}
with $\gr_K = \exp ( - c_\gb K^\gga)$.
\end{thm}

\noindent
\begin{rem}For the next results (Propositions \ref{Wulff compactness}, \ref{Wulff lower bound}, \ref{Wulff upper bound}) to hold true the mesoscopic scale $K$ has to be chosen just large enough (depending on $\b$ and on some extra parameter $\gd$). However in the next sections it will be essential to relate $K$ with the basic scale $N$ via the scaling relation $K \approx (\log N )^{1 / \gga}$ and therefore we will adopt this choice right away and denote the corresponding mesoscopic phase labels simply by $u_N$. Moreover, since the blocks with label $0$ will play an important role in the proof of the main results, they will be referred to as the {\it bad blocks}. \end{rem} In order to state the other results on the large deviations of $\mu_N^+$ we need to introduce some more notation.  For any $\gd >0$, the $\gd$-neighborhood of $v \in \bbL^1( \uTor)$ is defined by \begin{equation*}
\cV(v,\gd) = \{
v' \in \bbL^1( \uTor) \  |
\quad  \| v' - v  \|_1 < \gd \} \, .
\end{equation*}

Let $\cO$ be an open set containing $\uTor$.
The set of functions of bounded variation in $\cO$ taking values
in $\{-1,1\}$ and uniformly equal to 1 outside $\uTor$
is denoted by $\BV$ (see \cite{EG} for a review). For a
given $a >0$, the set of functions in $\BV$ with perimeter smaller than $a$ is denoted by $\cC_a$. Finally we define the Wulff functional $\cW_\gb$ on $\BV$ as follows. For any $v \in \BV$, there exists a generalized notion of the boundary of the set $\{ v = - 1 \}$ called reduced boundary and denoted by $\partial^* v$.  If $\{ v = - 1 \}$ is a regular set, then $\partial^* v$ coincides with the usual boundary $\partial v$. Then one defines \begin{equation*}
\cW_\gb(v) := \int_{\partial^* v} \st (\vec{n_x}) \, d \cH_x \, , \end{equation*} where $\cH_x$ is the $d-1$ Hausdorff measure. The Wulff functional $\cW_\gb$ can be extended on $\bbL^1(\uTor)$ by setting \begin{eqnarray} \label{functional F} \cW_\gb (v) = \left\{ \begin{array}{l}
\int_{\partial^* v} \st (\vec{n_x}) \, d \cH_x,
\qquad  {\rm if} \quad v \in \BV ,\\
\infty \; , \qquad \qquad \qquad \qquad  {\rm otherwise}. \end{array} \right. \end{eqnarray}
For any $m$ in $[-m^*,m^*[$,
the Wulff variational problem can then be stated as, \begin{eqnarray} \label{variational} \min \left \{ \cW_\gb (v) \ \Big| \ v \in \BV, \qquad
\big| \, \int_{\uTor} m^* \, v_r \, dr \big|  \leq m
\right\}.
\end{eqnarray}
If we denote by $\cD_m$ the set of minimizers of \eqref{variational} it has been proven by \cite{Taylor} that in $\bbR^d$
 the minimizer is unique up to
translations and given by suitable dilation of the Wulff shape \eqref{Wulff shape}.
In particular the interfacial energy of $\uTor$ is given by \begin{equation}
\label{eq: energy Wulff}
\tau^* = \cW_\gb (\uTor)=
\int_{\partial \uTor} \tau( \vec{n}_x ) d \cH_x \, . \end{equation} All that being said the results we are going to use can be summarized as follows.

\begin{pro}
\label{Wulff compactness}
There exists a constant $C(\gb)>0$ such that for any $\gd >0$ \begin{eqnarray*} \forall a >0, \qquad
\limsup_{N \to \infty} \;   \frac{1}{N^{d-1}} \log
\mu^+_N
\left( u_N \not \in \cV(\cC_a,\gd) \right)
\leq - C(\gb) \, a,
\end{eqnarray*}
where $\cV(\cC_a,\gd)$ is the $\gd$-neighborhood of $\cC_a$ in $\bbL^1( \uTor )$. \end{pro} \noindent This proposition tells us that only the configurations close to the compact set $\cC_a$ have a contribution which is of a surface order.\\ \noindent The precise asymptotic related to surface tension are \begin{pro} \label{Wulff lower bound} Uniformly over $\gd >0$ \begin{eqnarray*}
\liminf_{N \to \infty} \;   \frac{1}{N^{d-1}}
\log \mu^+_N
\big( \| u_N + 1 \|_1 \leq \gd \big)
\geq - \tau^*  \, .
\end{eqnarray*}
\end{pro}

\begin{pro}
\label{Wulff upper bound}
 For all $v$ in $\BV$ such that
$\cW_\gb(v)$ is finite and for $\gd >0$
\begin{eqnarray*}
\limsup_{N \to \infty} \; \frac{1}{N^{d-1}}
\log \mu^+_N
\big( \| u_N - v  \|_1 \leq \gd \big)
\leq - \cW_\gb(v) + \gep(\gd)  \, ,
\end{eqnarray*}
where $\gep(\gd)$ vanishes as $\gd$ goes to 0.
\end{pro}

\section{The test function}

In this section we define the test function that, if plugged into the Poincar{\'e} and logarithmic Sobolev inequalities, will prove theorems (\ref{pro spectral gap}) and (\ref{pro log sob}). As we mentioned in the introduction, the form of function described below was suggested by H.T. Yau.

Fix $\gl \in \, ]\frac{1}{2} \tau^* ,  \tau^* [$, where $\tau^*$ is defined in \eqref{eq: energy Wulff}. Let $g$ be a smooth non increasing function such that \begin{eqnarray*}
g(s) =
\begin{cases}
1, \qquad & \text{if} \quad   s \leq \frac{-m^*}{2} \, ,\\
0, \qquad & \text{if} \quad  s \geq \frac{- m^*}{4} \, . \end{cases} \end{eqnarray*} The mesoscopic scale $K$ is chosen equal to $(b \log N)^{\frac{1}{\gga}}$, where $\gga$ was introduced in
\eqref{eq domination} and $b$ is a constant which will be fixed later. The test function $f$ has the following expression: \begin{equation} \label{test function}
f(\gs) = \exp \left( \frac{\gl K^d}{N} \sum_j g(\cM_{N,K}(x_j)) \right) \, , \qquad \forall \gs \in \{ -1,1 \}^{\Tor{N}} \, . \end{equation} The factor $K^d$ stands for the volume of the boxes $\dBox{K}$ which equals to  $K^d$ (with the exception of some boxes along the boundary). Notice that $f$ is a non increasing function of the spins. \\ There are three main features of $f$ that make it quite effective. These are: \begin{enumerate}[\bf i)] \item The variance of $f$ almost coincides with $\mu_N^+(f^2)$ or, put it in another way, $\mu^+_N(f^2) \gg \mu_N^+(f)^2$; \item The entropy of $f^2$ w.r.t. to $\mu_N^+$ is of order $N^{d-1}$; \item Let us denote by $\mu_N^{+,f}$ the weighted measure $\frac{d \mu^{+,f}_N}{d \mu^+_N} = \frac{1}{Z^{+,f}_N} \, f^2$ where $Z_N^{+,f} := \mu_N^+(f^2)$. Then under $\mu_N^{+,f}$ the typical number of non zero terms in $|\nabla f|^2$ is  of the order of $N^{d-1}$. \end{enumerate}

\noindent
It is clear that once these properties are established then the proof of theorems \ref{pro spectral gap} and \ref{pro log sob} should follow quite easily.
\\
Intuitively the proof of ${\bf i}), {\bf ii})$ and ${\bf
iii})$ is based on the following simple heuristic. The
function $f$ assigns an exponential weight to the configurations with a large number of mesoscopic boxes with label $u_N=-1$ because of the choice of the function $g$. According to the large deviation theory, among the configurations favored by $f$, those with the largest $\mu_N^+$ weight form a Wulff droplet of a certain size. Therefore, to compute $\mu_N^+(f)$ or $\mu_N^+(f^2)$, we will need to compare, for a given Wulff droplet, the gain given by the exponential factor in $f$ or $f^2$ with the $\mu_N^+$ probability of creating the droplet itself. It turns out, due to the precise choice of the parameter $\l$, that the balance for $f$ is negative and no Wulff droplet will
appear, while the balance is positive for $f^2$ and the typical spin configurations under $\mu_N^{+,f}$ will consist of a Wulff droplet of the {\it minus phase} of volume $\approx N^d$. That accounts for {\bf
i}) and {\bf ii}).
Given the above picture, it is also clear that {\bf iii}) holds simply because the non zero terms in $|\nabla f|^2$ come only from the bad boxes, again because of the choice of the function $g$. The boundary of the Wulff droplet produces $O(N^{d-1})$ of such boxes while the inside of the droplet typically does not contain {\it any} bad box because of the choice of the mesoscopic scale $K$. Were $K$ be large but independent of $N$ then we would always have a density of bad boxes and the whole construction would break down.

\subsection{ The Variance of $f$}
\label{subsec: variance}

We are first going to check that
\begin{equation}
\label{variance 1}
\lim_{N \to \infty}
\frac{\mu^+_N (f,f)}{\mu^+_N (f^2)}
=
\lim_{N \to \infty}
\frac{\mu^+_N (f^2) - \mu^+_N (f)^2}{\mu^+_N (f^2)} = 1  \, . \end{equation} The function uniformly equal to $-1$ in $\uTor$ is denoted by $-\1$. Let $\gep >0$, then \begin{eqnarray*} \mu^+_N (f^2) \geq \mu^+_N \left( f^2 \,
1_{\{ u_N\in \cV( - \1, \gep) \}}
\right)
\geq \exp \left( \frac{2 \gl}{N} N^d (1-\gep) \right)
\mu^+_N \left( u_N\in \cV( - \1, \gep) \right) \, , \end{eqnarray*} where we used the fact that if $u_N (x) = -1$ then
$\cM_{N,K}(x) < - \frac{m^*}{2}$.
Proposition \ref{Wulff lower bound} implies that for $N$ large enough \begin{eqnarray} \label{eq gap 2} \mu^+_N (f^2) \geq \exp \left( N^{d-1}
\big( 2 \gl (1-\gep) - \tau^* - o(\gep) \big) \right) \, , \end{eqnarray} where $o(\gep)$ vanishes as $\gep$ goes to 0.\\
Next we examine
$\mu_N^+(f)$ and prove that
\begin{eqnarray}
\label{eq gap 3}
\lim_{N \to \infty} \frac{1}{N^{d-1}}
\log
\mu^+_N (f)
= 0 \, .
\end{eqnarray}
The derivation of an upper bound for $\mu^+_N (f)$ requires some technicalities.  First fix a constant $a > \frac{\gl}{C(\gb)}$ where $C(\b)$ appears in Proposition \ref{Wulff compactness}. Then Proposition \ref{Wulff compactness} implies that for any $\gd >0$ \begin{eqnarray} \label{variance 2.1} \limsup_{N \to \infty} \; \frac{1}{N^{d-1}} \log \mu^+_N
\left( f \,  1_{\{ u_N\not \in \cV( \cC_a , \gd) \} }
\right)
\leq \gl - C(\gb)\, a < 0  \, .
\end{eqnarray}
Then fix $\gep >0$ and recall that $\cC_a$ is compact for the $\bbL^1$ topology.  According to Proposition \ref{Wulff upper bound}, for $\gd$ small enough the set $\cV( \cC_a , \gd)$ can be covered by a finite union $\cup_{i = 1}^\ell \cV( v_i, \gep_i)$ such that for any $i \leq \ell$ and $N$ large enough \begin{eqnarray} \label{variance 2.2}  \frac{1}{N^{d-1}} \log
\mu^+_N \big( u_N\in \cV( v_i, \gep_i)
\big)
\leq
- \cW_\gb (v_i) + \gep \, ,
\end{eqnarray}
where $\gep_i \in (0,\gep)$. Noticing that
\begin{eqnarray}
\label{variance 2}
\mu^+_N (f) \leq
\sum_{i =1}^\ell
\mu^+_N \left(
f \,  1_{ \{ u_N\in \cV( v_i, \gep_i) \} }
\right)
+ \mu^+_N
\left(
f \,  1_{\{ u_N \not \in \cV( \cC_a , \gd) \} }
\right) \, ,
\end{eqnarray}
and combining \eqref{variance 2.1} with \eqref{variance 2.2}, we get \begin{eqnarray*} \mu^+_N (f) \leq
\sum_{i =1}^\ell
\exp \left( N^{d-1}
\big( \gl |v_i| - \cW_\gb (v_i) + \gep (1 +\gl) \big) \right)
+ \exp \left( N^{d-1}
\big( \gl - C(\gb) a  \big) \right) \, ,
\end{eqnarray*}
where $| v_i |$ denotes the volume of the set $\{ v_i = -1 \}$. To check that the spin configuration in $\{  u_N\in \cV( v_i, \gep_i)\}$ have a number of blocks of the order of $N^d |v_i|$, it is enough to regularize $v_i$ by a polyhedral set (see Thm 2.5.1 in \cite{BIV}). \\ By the very definition of the Wulff variational problem, for
any $v \in \BV$
\begin{eqnarray}
\label{Wulff variational}
\cW_\gb (v) \geq \tau^* |v|^{(d-1)/d}
\geq \tau^* |v| \, ,
\end{eqnarray}
where we have used the fact that $|v| \leq |\uTor| = 1$.
As $\gl < \tau^*$,
\begin{eqnarray}
\label{variance 3}
\mu^+_N (f) \leq \ell
\exp \left( N^{d-1}  \gep (1 +\gl) \right) + 1 \, . \end{eqnarray} Since $\gep$ was arbitrary, this implies \eqref{eq gap 3}.\\ Combining \eqref{eq gap 2}, \eqref{eq gap 3} and the fact that $2 \gl > \tau^*$, we finally derive \eqref{variance 1} by choosing $\gep$ small enough.

\subsection{ The Entropy of $f^2$}

We will prove that
\begin{equation}
\label{eq LS 1}
C\,N^{d-1} \leq \mu^{+,f}_N (\log f^2 ) - \log \mu^+_N (f^2)  \, , \end{equation} for a suitable constant $C$.\\ Using the previous strategy, we check that for
$\gep>0$ and for $N$ large enough
\begin{eqnarray*}
\mu^+_N (f^2) \leq
\sum_{i =1}^\ell
\exp \left( N^{d-1}
\big( 2 \gl |v_i| - \cW_\gb (v_i) + \gep (1 +\gl) \big) \right)
+ 1 \, .
\end{eqnarray*}
Inequality \eqref{Wulff variational} implies
\begin{eqnarray}
\label{LS 3}
\nonumber
\mu^+_N (f^2) & \leq &
\sum_{i =1}^\ell
\exp \left( N^{d-1}
\big( (2 \gl - \tau^*) |v_i|  + \gep (1 +\gl) \big) \right)
+ 1 \, ,\\
& \leq &
\ell \exp \left( N^{d-1}
\big( 2 \gl - \tau^*  + \gep (1 +\gl) \big) \right)
+ 1 \, .
\end{eqnarray}
It remains to check that for $\gep >0$ and $N$ large enough \begin{eqnarray} \label{LS 4} \mu^{+,f}_N \big( \log f^2 \big) = \frac{2 \gl K^d}{N} \mu^{+,f}_N
\Bigl(
\sum_{x \in \sTor{N,K}}g(\cM_{N,K}(x))
\Bigr)  \geq (1 - \gep) 2 \gl N^{d-1} \, .
\end{eqnarray}
This is a consequence of the following estimate.
For any $\gep>0$ and $N$ large enough
\begin{eqnarray}
\label{LS 5}
(1-o(\gep)) \mu^{+}_N (f^2)
\leq
\mu^{+}_N \left(
f^2 1_{\{ u_N \in \cV( - \1, \gep)\}}
\right) \, .
\end{eqnarray}
Let $\cF = \big( \cV( - \1, \gep) \big)^c$. First notice that \begin{eqnarray*} \sup_{v \in \cF} \left\{
- \cW (v) + 2 \gl |v| \right\}
\leq
\sup_{v \in \cF} \left\{ |v| (
- \tau^* + 2 \gl) \right\}
\leq (- \tau^* + 2 \gl) (1 - \gep) \, .
\end{eqnarray*}
We proceed as before and cover the set $\cF \cap
\cV (\cC_a, \gd) $ with a finite number of neighborhoods.
This implies that for any $\gd >0$ and $N$ large enough \begin{eqnarray*} \mu^{+}_N \left( f^2 1_{\{ u_N  \in \cF \}}
\right)
\leq \ell
\exp \big( (-\tau^* + 2 \gl + o(\gd))(1 - \gep)
N^{d-1}
\big) + 1 \, .
\end{eqnarray*}
On the other hand,
\begin{eqnarray*}
\mu^{+}_N \Bigl(
f^2 1_{\{ u_N \in \cV( - \1, \gep)\}}
\Bigr)
\geq
\exp \big( (-\tau^* + 2 \gl) N^{d-1}
\big) \, .
\end{eqnarray*}
Thus, for $N$ large enough, we derive \eqref{LS 5}.
This implies that
\begin{eqnarray*}
K^d \mu^{+}_N
\Bigl( f^2 \, (
\sum_{x \in \sTor{N,K}}g(\cM_{N,K}(x))
)
\Bigr)  \geq
\mu^{+}_N
\left( f^2 \, (
1_{\{ u_N \in \cV( - \1, \gep)\}}
)
\right)
(1 - \gep)  N^d \, .
\end{eqnarray*}
The inequality \eqref{LS 5} leads to
\begin{eqnarray*}
K^d \mu^{+}_N
\Bigl( f^2 \, (
\sum_{x \in \sTor{N,K}}g(\cM_{N,K}(x)) )\Bigr)  \geq
\mu^{+}_N
\left( f^2 \right)
(1 - \gep) (1 - o(\gep))  N^d \, .
\end{eqnarray*}
Therefore \eqref{LS 4} is complete.
Combining \eqref{LS 3} and \eqref{LS 4}, we obtain
for $N$ large enough
\begin{eqnarray*}
\mu^{+,f}_N (\log f^2 ) - \log \mu^+_N (f^2)
&\geq& \big(
(1-o(\gep)) 2 \gl - ( -\tau^* + 2\gl + \gep^\prime (1 +\gl))
\big) N^{d-1} - \log \ell \, , \\
& \geq &
\left(\tau^*
-o(\gep) 2 \gl  - \gep^\prime (1 +\gl)
\right) N^{d-1} - \log \ell \, .
\end{eqnarray*}
For any $\gb > \gb_c$ (the true critical point) it is known that $\tau^*>0$. Thus, by choosing $\gep$ and $\gep'$ small enough, we derive
\eqref{eq LS 1} for $N$ large enough.

\subsection{ The Dirichlet Form of $f$}

The Dirichlet form associated to $f$ can be bounded as follows. There is $C_1>0$ such that for $N$ large enough \begin{equation} \label{dirichlet} \mu^+_N \big( | \nabla f |^2 \big) \leq
C_1 \gl^2 N^{d-3} K^d \;  \mu^+_N \big( f^2 \big)
=
C_1 \gl^2 N^{d-3} (b \log N)^{\frac{d}{\gga}}   \mu^+_N \big( f^2 \big)
\, .
\end{equation}
By Taylor expansion
\begin{eqnarray*}
| \nabla f |^2 =
\sum_{i \in \Tor{N}}  | \nabla_i f |^2
&=&
\sum_{x \in \sTor{N,K}}
\sum_{i \in \dBox{K}(x)} | \nabla_i f |^2\\
&\leq & f^2\,\Bigl(\,2 K^d  \frac{\gl^2}{N^2} \| g' \|_\infty^2\,\Bigr) \sum_{x \in \sTor{N,K}}
 1_{\{  - \frac{m^*}{2} \leq \cM_{N,K}(x) \leq - \frac{m^*}{4} \}} \\ &\leq & f^2\,\Bigl(\,2 K^d  \frac{\gl^2}{N^2} \| g' \|_\infty^2\,\Bigr) \cQ_N \, . \end{eqnarray*} where $\cQ_N$ denotes the number of blocks in $\Tor{N}$ with averaged magnetization in $[-\frac{m^*}{2}, - \frac{m^*}{4}]$.
Using the notation $\mu^{+,f}_N$ introduced in {\bf ii}) above we can
write
\begin{equation*}
\mu^+_N \big( | \nabla f |^2 \big)
\leq
c_1 \frac{\gl^2 K^d}{N^2} \| g' \|_\infty^2 \,
\mu^+_N ( f^2 ) \;
\mu^{+,f}_N ( \cQ_N )
\, .
\end{equation*}
The estimate \eqref{dirichlet} will follow from the fact that for $N$ large enough \begin{equation} \label{dirichlet 1} \mu^{+,f}_N \big( \cQ_N \big) \leq 2 N^{d-1} \, . \end{equation} This boils down to check that \begin{equation} \label{dirichlet 2} \mu^{+,f}_N \left( \cQ_N  \, 1_{\{ \cQ_N > N^{d-1} \}} \right) \leq N^d \exp ( - c N^{d-1})\, , \end{equation} where $c$ is a positive constant. As $\mu^{+}_N (f^2) \geq 1$,  we see that \begin{equation} \label{dirichlet 3} \mu^{+,f}_N \left( \cQ_N  \, 1_{\{ \cQ_N > N^{d-1} \}} \right) \leq \exp ( 2 \gl N^{d-1})
\mu^{+}_N \left( \cQ_N > N^{d-1} \right)
\frac{N^d}{K^d} \, .
\end{equation}
Remember that the occurrence of bad blocks is dominated by Bernoulli percolation with parameter $\gr_K = N^{- c_\gb b}$.
Therefore, for $b$ large enough, it is then quite simple to check that there is $c>0$ such that for  $N$ large enough, \begin{equation*} \mu^{+}_N \left( \cQ_N > N^{d-1}  \right) \leq \exp \big(  - c N^{d-1} \big) \, . \end{equation*} Combining the previous bound with \eqref{dirichlet 3}, we derive \eqref{dirichlet 2}. \begin{rem} It would be possible  to derive sharper estimates for \eqref{dirichlet
  1}. One expects
\begin{equation*}
\mu^{+,f}_N \big( \cQ_N \big) \leq c \frac{N^{d-1}}{K^{d-1}} \, . \end{equation*} Nevertheless this would not be enough to derive an asymptotic for the spectral gap and the Log-Sobolev constant without a logarithmic correction : on finite mesoscopic scales, we cannot control the test function. \end{rem}

\vfill\eject
\section{ Proof of Theorems \ref{pro spectral gap} and \ref{pro log sob}} \noindent We are in position to prove the first two main results.
\subsection{Proof of Theorem \ref{pro spectral gap}} By definition
$$
 S_N \ge \frac{\mu_N^+(\phi,\phi)}{\cE(\phi,\phi)},  \qquad \forall \phi
$$
When $\phi$ is equal to our test function $f$ the above ratio can be bounded from below using \eqref{dirichlet} and \eqref{variance 1} by \begin{equation} \label{bound spectral gap}
\frac{N^{3-d}}{ C_2  (  \log N)^{d/\gga}} \, ,
\end{equation}
where $C_2 = C_1 \gk^{d / \gga} \gl^2$. \qed
\\

\noindent
Clearly in dimension $d \geq 3$
the test function does not provide any information.

\subsection{Proof of Theorem \ref{pro log sob}}

Fix $d \geq 2$. By definition
$$
L_N \geq \frac{\mu_N^+(\phi^2\log \phi^2)}{\cE_N^+(\phi,\phi)},
\qquad  \forall \, \phi
$$
When $\phi$ is equal to our test function $f$ the above ratio can be bounded from below using \eqref{dirichlet} and \eqref{eq LS 1} by \begin{equation} \label{bound log sob}
\frac{N^{2}}{ C_3  (  \log N)^{d/\gga}} \,
\end{equation}
where $C_3 = \frac{C}{C_1 \gk^{d / \gga} \gl^2}$. \qed

\vskip 1cm
\section{Slow down of the Glauber dynamics in two dimensions}
\label{sec: Glauber dynamics}

\noindent
In this section we will derive some consequences from the two dimensional upper bound on the inverse spectral gap for $\b > \b_c$ on the speed of relaxation of the Glauber dynamics to its equilibrium. In particular we will prove theorem \ref{stretch decay}. The notation will be that fixed in section 2 and section 3.

\subsection{A first finite volume bound}
The first simple consequence of Theorem \ref{pro spectral gap}, is a bound on the dynamical evolution of the test function \eqref{test function} itself.

\begin{pro}
\label{thm: decay finite volume}
For any $N$ large enough,
\begin{eqnarray}
\label{eq: decay finite volume}
\forall t> 0 , \qquad
\mu^+_N \left( \big( P_t^{(+,N)} f \big)^2 \right)
\geq  \mu^+_N (f^2) \exp \left( - 2t \frac{(\log N)^\gk}{N} \right)
\big( 1 - \exp(-c_\gl N) \big) \, ,
\end{eqnarray}
where $c_\gl$ is a positive constant depending on $\gl$. \end{pro}

This result provides a first (admittedly weak) clue on the relaxation time of the dynamics. Let us assume that the Markov process generated by $L_N^+$ is attractive (see \cite{Li} or \cite{M1}). This is the case if for example the flip rates were those of the Metropolis or of the Heat Bath dynamics. Let $\cB^-_N(\gs)$ be the number blocks $\dBox{K}$ for which the spin configuration $\gs$ in $\{ \pm 1\}^{\Tor{N}}$ has averaged magnetization smaller than $- \frac{m^*}{4}$. We set \begin{eqnarray*} \Psi_N (\gs) = \exp \left( \frac{\gl \, K^d}{N} \cB_N^- (\gs) \right) \, , \end{eqnarray*} where $K = (b\log N)^{1/\gga}$ and $b,\g$ are as in the previous section.\\ Since $\cB^-_N$ is a non increasing function of the spin variables, the monotonicity inequalities for attractive processes imply \begin{eqnarray*} P_t^{(+,N)} \big( \Psi_N  \big)(-) \geq \left( \mu^+_N \left( \big( P_t^{(+,N)} \Psi_N \big)^2 \right) \right)^{1/2} \geq \left( \mu^+_N \left( \big( P_t^{(+,N)} f \big)^2 \right) \right)^{1/2} \, , \end{eqnarray*} where the symbol $(-)$ denotes the configuration in $\Tor{N}$ for which all the spins are equal to $-1$. Inequality \eqref{eq: decay finite volume} implies that there is $\gep > 0$ such that for all $N$ large enough \begin{eqnarray*} P_t^{(+,N)} \big( \Psi_N  \big)(-) \geq  \exp \left( (2 \gl - \tau^* - \gep) \frac{N}{2}
- t \frac{(\log N)^\gk}{N} \right) \, .
\end{eqnarray*}
On the other hand, as in the derivation of \eqref{eq gap 3}, one can check that \begin{eqnarray*} \lim_{N \to \infty} \frac{1}{N} \log \mu^+_N \big( \Psi_N \big) = 0 \, . \end{eqnarray*} Therefore for any time smaller than $\frac{N^2 }{(\log N)^{ 2\gk}}$ the quantity $P_t^{(+,N)} \big( \Psi_N (\gs) \big)(-)$ is much larger than the equilibrium expectation of $\Psi_N$ : in the above special sense the system has not yet relaxed. \begin{rem} It is important to observe that in the above reasoning we have never used the information that the logarithmic Sobolev constant is larger than $\approx N^2$. Unfortunately we have not been able to establish anything like proposition \ref{thm: decay finite volume} for the entropy of $\bigl(\,P_t^{(+,N)}f\,\bigr)^2$ with the exponent $t \frac{(\log N)^\gk}{N}$ replaced by $t \frac{(\log N)^\gk}{N^2}$. \end{rem}

\begin{proof}
We set $\phi = f -  \mu^+_N(f)$.
The spectral decomposition of $\cL^+_N$ implies \begin{eqnarray*} \mu^+_N \left( \big( P_t^{(+,N)} \phi \big)^2 \right) =  \int_0^\infty \, d\nu_\phi (\gt) \exp ( - 2t \gt) \, , \end{eqnarray*} where $\nu_\phi$ denotes the spectral measure associated to $\phi$. By Jensen inequality, \begin{eqnarray*} \mu^+_N \left( \big( P_t^{(+,N)} \phi \big)^2 \right) \geq
\left( \int_0^\infty \, d\nu_\phi (\gt) \right) \;
\exp \left( - 2t \frac{ \int_0^\infty  \gt \, d\nu_\phi (\gt)  } {\int_0^\infty \, d\nu_\phi (\gt)}
\right) \, .
\end{eqnarray*}
By definition of the spectral measure
\begin{eqnarray*}
\mu^+_N \big( \phi^2 \big) =
\int_0^\infty \, d\nu_\phi (\gt)
\quad \text{and} \quad
\cE_N^+(f,f)\, =\,
\int_0^\infty  \gt \, d\nu_\phi (\gt)
\; .
\end{eqnarray*}
Therefore the bound on the spectral gap
(see \eqref{bound spectral gap}) implies that for $N$ large enough,
\begin{eqnarray}
\label{eq: decay finite volume 2}
\mu^+_N \left( \big( P_t^{(+,N)} \phi \big)^2 \right)
\geq  \mu^+_N \big( \phi^2 \big) \;
\exp \left( - 2t \frac{(\log N)^\gk}{N} \right) \, . \end{eqnarray} According to \eqref{eq gap 3}, there is a constant $c_\gl > 0$ such that
\begin{eqnarray*}
\mu^+_N ( f )^2 \leq \mu^+_N ( f^2 ) \exp( - c_\gl N) \, . \end{eqnarray*} The former inequality combined with \eqref{eq: decay finite volume 2} leads to
\begin{eqnarray}
\mu^+_N \left( \big( P_t^{(+,N)} f \big)^2 \right)
\geq  \mu^+_N \big( f^2 \big) \;
\exp \left( - 2t \frac{(\log N)^\gk}{N} \right)
\big( 1 - \exp(- c_\gl N) \big) \, .
\end{eqnarray}
This concludes the proof.
\end{proof}

\subsection{Proof of Theorem \ref{stretch decay}}

\begin{proof}
The first step is to reformulate the LHS of \eqref{eq: stretch decay} in terms of mesoscopic variables. For any site $x \in K\,\bbZ^2$ we define $\gz^\eta_x (t)$ to be the indicator function of the event that the magnetization in the box $\dBox{K}(x)$ for the process $\gs^\eta (t)$ in $\dBox{K}(x)$ is smaller than $- \frac{m^*}{4}$. Then we have \begin{eqnarray*} {\hat \mu}^+ \left( {\hat \bbE} \big( \gz_0^\eta (t) \neq {\tilde \gz}_0^\go (t)
\big) \right)
&\leq& {\hat \mu}^+ \left(
{\hat \bbE} \big(\, \exists \, i \in \dBox{K}(0), \quad \gs_i^\eta (t) \neq {\tilde \gs}_i^\go (t)
\,\big)\right) \\
&\leq&
K^2 {\hat \mu}^+ \left(
{\hat \bbE} \big(\, \gs_0^\eta (t) \neq {\tilde \gs}_0^\go (t)
\,\big) \right) \, ,
\end{eqnarray*}
where we used the invariance by spatial translation in the last
inequality.

\noindent
Let $N$ be a large integer, choose as usual the mesoscopic scale $K = (b\log N)^{1/\gga}$ and let $L = \frac{N}{K}$.  By repeating the previous computation on a coarse grained level, we get \begin{eqnarray*} {\hat \mu}^+ \Bigl( {\hat \bbE} \big( \sum_{i \in \Tor{N} \cap K \bbZ^2} \gz_i^\eta (t) \neq \sum_{i \in \Tor{N} \cap K \bbZ^2}
{\tilde \gz}_i^\go (t)
\big) \Bigr)
&\leq&
{\hat \mu}^+ \left( {\hat \bbE}
\big( \exists\, i \in  \Tor{N} \cap K \bbZ^2, \quad
\gz_i^\eta (t) \neq {\tilde \gz}_i^\go (t)
\big)
\right)\\
&\leq&
L^2 {\hat \mu}^+ \left(
{\hat \bbE} \big( \gz_0^\eta (t) \neq {\tilde \gz}_0^\go (t)
\big)
\right) \, .
\end{eqnarray*}
Let now
\begin{eqnarray*}
\cB^-_N (\gs^\eta_t) = \sum_{i \in \Tor{N} \cap K\bbZ^2} \gz_i^\eta (t) \, . \end{eqnarray*} The previous results imply \begin{eqnarray}
\label{eq: stretch 1}
{\hat \mu^+_N} \left(
{\hat \bbE} \big( \gs_0^\eta (t) \neq {\tilde \gs}_0^\go (t)
\big) \right)
\geq \frac{1}{N^2}
{\hat \mu}^+ \left(
{\hat \bbE} \big( \cB^-_N(\gs^\eta_t) \neq \cB^-_N ({\tilde \gs}^\go_t)
\big) \right)  \, .
\end{eqnarray}

\noindent
In the second step, we are going to decouple the estimates of the joint process. The main physical idea was already contained in the Fisher, Huse paper \cite{HF} and it goes as follows. We force one large droplet of the minus phase of radius $\approx N$, around the origin in e.g. the initial distribution of $\s^\h_t$, by paying a price $\approx
\exp(-\tau^* \,N)$. This droplet should relax only in a time scale proportional to its initial area and therefore, if $N= A\sqrt{t}$ with $A$ large enough, the distribution of $\cB^-_N(\gs^\eta_t)$ at time $t$ given the above initial unlikely event should be quite different from that of $\cB^-_N(\gs^\o_t)$. Apparently in order to carry rigorously the above program one needs a much more precise control on the life time of a droplet than what we have been able to obtain. Actually that is not true and all what we need is something not more precise than proposition
\ref{thm: decay finite volume} (see Lemma \ref{lemme stretch decay} below). \\ From a technical point of view it is convenient to force the droplet of the minus phase inside $\mu^+(\h)$ in a ``soft'' way by simply inserting our test function $f^2$ defined in \eqref{test function} with $N \approx \sqrt{t}$. \\ We write \begin{gather}
\label{eq: joint process}
{\hat \mu}^+ \big(
{\hat \bbE} \big( \cB^-_N(\gs^\eta_t)  \neq \cB^-_N ({\tilde \gs}^\go_t)
\big) \big)
\geq \nonumber \\
\geq \mu^{+}(f^2) \exp( - 2 \gl N)
\int \, d \mu^{+,f^2} (\eta) \, d\mu^+ (\go) \,
{\hat \bbE} \big( \cB^-_N(\gs^\eta_t)  \neq \cB^-_N ({\tilde \gs}^\go_t)
\big)  \geq \nonumber \\
\geq \exp( - 2 \t^* N)
\int \, d \mu^{+,f^2} (\eta) \, d\mu^+ (\go) \,
{\hat \bbE} \big( \cB^-_N(\gs^\eta_t)  \neq \cB^-_N ({\tilde \gs}^\go_t)
\big)
\end{gather}
Let $\ga$ be a parameter in $(0,1)$ which will be fixed
later on. Then
\begin{eqnarray*}
{\hat \bbE} \big( \cB^-_N(\gs^\eta_t) \neq \cB^-_N ({\tilde \gs}^\go_t)
\big)
&\geq&
{\hat \bbE} \big( \cB^-_N(\gs^\eta_t)  > \ga L^2 \; ;
\cB^-_N ({\tilde \gs}^\go_t)  \leq \ga L^2
\big) \\
&\geq& \bbE \big( \cB^-_N(\gs^\eta_t) > \ga L^2 \big) -
\bbE \big( \cB^-_N ({\tilde \gs}^\go_t) > \ga L^2 \big)  \, , \end{eqnarray*} where $\bbE$ refers to the marginal of ${\hat \bbE}$, i.e. to the usual Glauber dynamics. Since the measure $\mu^+$ is invariant with respect to the Glauber
dynamics, we can write
\begin{eqnarray*}
\mu^+ \left(
\bbE \big( \cB^-_N(\gs^\eta_t)  > \ga L^2 \big)
\right)
= \mu^+  \big( \cB^-_N(\gs)  > \ga L^2 \big)  \leq \exp( - C_\ga N)
\, ,
\end{eqnarray*}
where the final estimate follows from the theory of equilibrium phase coexistence (see Propositions \ref{Wulff upper bound} and \ref{Wulff compactness}).  In conclusion \begin{eqnarray*} {\hat \mu}^+ \left( {\hat \bbE} \big( \cB^-_N(\gs^\eta_t)  \neq \cB^-_N ({\tilde \gs}^\go_t)
\big) \right)
\geq
\exp( - 2\tau^*   N)
\left( \mu^{+,f^2} \left(
\bbE \big( \cB^-_N(\gs^\eta_t)  > \ga L^2 \big) \right)
 - \exp( - C_\ga N)  \right) \, .
\end{eqnarray*}
It is at this stage that we are going to use the information on the spectral gap.
The necessary dynamical estimate is provided by the following Lemma  which will be derived later. \begin{lem} \label{lemme stretch decay} We fix  $\ga$ such that the parameter $(2\gl (\ga - 1) + \tau^*)$ is negative.
Then, for $N$ large enough, the following inequality holds
\begin{eqnarray*}
\forall t >0, \qquad
\int \, d \mu^{+,f^2} (\eta) \;
\bbE \big( \cB^-_N( \gs^\eta_t) > \ga L^2 \big)
\geq \frac{1}{2} \exp\Bigl(-t\frac{(\log N)^\gk}{N}\Bigr) -
\exp( -c_{\ga,\gl} N)
 \, .
\end{eqnarray*}
where $c_{\ga,\gl} >0$.
We recall that $L = \frac{N}{K}$.
\end{lem}

\noindent The previous Lemma implies that
\begin{eqnarray*}
&&{\hat \mu}^+ \left(
{\hat \bbE} \big( \cB^-_N(\gs^\eta_t)  \neq \cB^-_N ({\tilde \gs}^\go_t)
\big) \right)
\geq \\
&& \qquad \qquad \qquad
\exp\Bigl( - 2\tau^* N \Bigr)
\left( \frac{1}{2} \exp\Bigl( - t \frac{(\log N)^\gk}{N} \Bigr) -
\exp( -c_{\ga,\gl} N)  - \exp( - C_\ga N)  \right) \, . \end{eqnarray*} By choosing $ N = \sqrt{t}\, (\log t)^\gk$, we finally derive for $t$ large enough \begin{eqnarray*} {\hat \mu}^+ \left( {\hat \bbE} \big( \cB^-_N(\gs^\eta_t)  \neq \cB^-_N ({\tilde \gs}^\go_t)
\big) \right)
\geq
\exp\Bigl(-2\tau^*\sqrt{t} (\log t)^\gk
 - 3 \sqrt{t} \Bigr) \, .
\end{eqnarray*}
\end{proof}

\vskip0.5cm
\noindent
{\it Proof of Lemma \ref{lemme stretch decay}.}
The proof relies on the dynamical estimate of proposition
\ref{thm: decay finite volume}.
\\Let
\begin{eqnarray}
\label{lem5 eq1}
\Psi_t = \mu^+\Bigl( \big( P_t f \big)^2 \Bigr)
=
\mu^+\left( \left[
\bbE \big( f(\gs^\eta_t) \, 1_{\cB^-_N(\gs^\eta_t) < \ga L^2} \big)
+ \bbE \big( f(\gs^\eta_t) \, 1_{\cB^-_N(\gs^\eta_t) \geq \ga L^2} \big)
\right]^2 \right) \, .
\end{eqnarray}
Thus from the estimate \eqref{eq gap 2} and the FKG inequality we see that for $\gep$ small enough, $\ga$ such that
$(2\gl (\ga - 1) + \tau^*)\equiv \d_\a <0$ and $N$ large \begin{gather*} \bbE \big( f (\gs^\eta_t)\, 1_{\cB^-_N (\gs^\eta_t) < \ga L^2} \big) \leq \exp( \ga \gl N ) = \exp \left( (2 \gl - \tau^* - \d_\a) \frac{N}{2} \right) \leq \\ \leq \exp(-\frac{\d_\a}{2}N)\sqrt{\mu_N^{+} (f^2)} \leq
\exp(-\frac{\d_\a}{2}N) \sqrt{\mu^{+} (f^2)} \, .
\end{gather*}
Plugging the above inequality in \eqref{lem5 eq1}, we get \begin{eqnarray}
\label{eq: blocks decay}
\Psi_t \leq
2 \,  \mu^+ \left( \left[
 \bbE \big( f(\gs^\eta_t) \, 1_{\cB^-_N(\gs^\eta_t) \geq \ga L^2} \big)
\right]^2   \right)
+
\mu^{+} \big(f^2  \big) \exp(-\frac{\d_\a}{2}N)  \, , \end{eqnarray} with $\d_\ga >0$. \\ In the pure phase $\mu^+$, the estimates obtained in subsection
\ref{subsec: variance} for the variance and the Dirichlet
form of $f$ hold (see Proposition \ref{infinite volume} below). Thus Proposition \ref{thm: decay finite volume} is also valid for an unbounded region and for $N$ large enough, we get \begin{eqnarray*} \forall t >0, \qquad \Psi_t \geq  \mu^{+} (f^2) \exp \left( - 2 t \frac{ (\log N)^\kappa}{N} \right)
\big( 1 - \exp ( - c_\gl N)  \big) \, .
\end{eqnarray*}

Combining the previous inequality with \eqref{eq: blocks decay}, we get by using Cauchy Schwartz inequality \begin{eqnarray*} \mu^{+} \left(  \bbE \big( f^2(\gs^\eta_t) \big) \, \bbE \big(
  1_{\cB^-_N(\gs^\eta_t) \geq \ga L^2} \big)  \right)
\geq  \mu^{+} (f^2) \left( \frac{1}{4}
\exp \left( - 2 t \frac{ (\log N)^\kappa}{N} \right)
- \exp( -c_\ga N)
\right) \, .
\end{eqnarray*}
The reversibility of the dynamics ensures that \begin{eqnarray*} \mu^{+} \left(  \bbE \big( f^2(\gs^\eta_t) \big) \, \bbE \big(
  1_{\cB^-_N(\gs^\eta_t) \geq \ga L^2} \big)  \right)
=
\mu^{+} \left(  f^2(\eta)  \, \bbE \big(
  1_{\cB^-_N(\gs^\eta_{2t}) \geq \ga L^2} \big)  \right)  \, . \end{eqnarray*} This concludes the Lemma.
\qed

\medskip

\begin{pro}
\label{infinite volume}
In dimension $d=2$, for any $\gb > \gb_c$ then
$$
\forall N, \qquad
\mu^+(f^2) - \mu^+ (f)^2 \geq C \frac{N}{(\log N)^\gk}
\mu^+(| \nabla f |^2) \, ,
$$
where $f$ is the test function introduced in \eqref{test function}. \end{pro}

\begin{proof}
The upper bound \eqref{dirichlet} on the Dirichlet form is unchanged
under $\mu^+$ since it boils down to estimating the number of bad blocks in the region $\Tor{N}$ by using Bernoulli percolation. The lower bound \eqref{eq gap 2} holds also for $\mu^+ (f^2)$ because it involves only the computation of an event localized in
$\Tor{N}$.

Thus it remains only to check that
\begin{eqnarray}
\label{infinite vol 1}
\limsup_{N \to \infty} \; \frac{1}{N} \log \mu^+ (f)
\leq 0 \, .
\end{eqnarray}

Let $\widehat D = [- R, R]^2 \subset \bbR^2$, where $R$ will be chosen large
enough; in particular such that ${\widehat \bbW}^{\, 2} \subset [- R/2, R/2]^2$. Let $\cC_N$ be the set of spin configurations which contain a $*$--connected
circuit of $+$
spins inside $[-N,N]^2 \setminus [-N/2,N/2]^2$ separating $([-N,N]^2)^c$ from
$[-N/2,N/2]^2$.
As $\gb > \gb_c$ and $d=2$, there is $c_\gb >0$ such that \begin{eqnarray*} \mu^+ (\cC_N^c) \leq \exp(-c_\gb N) \, .
\end{eqnarray*}
By choosing $R$ such that $R c_\gb > 2 \gl$, we get \begin{eqnarray*} \mu^+ (f) \leq \mu^+ (f \, 1_{\cC_{NR}}) + \mu^+ ( \cC_{NR}^c )\, \exp(\gl N)
\leq \mu^+ (f \, 1_{\cC_{NR}})+ \exp( (\gl - c_\gb R) N) \, .
\end{eqnarray*}
Conditionning with respect to the $+$ circuit which is the closest to $([-RN,RN]^d)^c$ and then using the fact that $f$ is non-increasing, we obtain by FKG $$ \mu^+ (f 1_{\cC_{NR}}) \leq \mu^+_{NR} (f ) \, .
$$
where $\mu^+_{NR}$ denotes now the Gibbs measure in $\{ - NR, \dots,  NR\}^2$ with $+$ boundary conditions.

At this point, we proceed as in subsection \ref{subsec: variance}. The only difference is that the estimates are in $\bbL^1(\widehat D)$ instead of $\bbL^1({\widehat \bbW}^2)$.
This implies
\begin{eqnarray*}
\limsup_{N \to \infty} \; \frac{1}{N} \log \mu^+_{NR} (f)
\leq \sup_{v \in \text{BV}(\widehat D, \{ \pm 1\})}
\Big\{ - \cW_\gb(v) + \gl \big| \{ v =-1\} \cap {\widehat \bbW}^2 \big| \Big\} \,. \end{eqnarray*} To derive \eqref{infinite vol 1}, it remains to check that the RHS is negative. Either $|v| \geq 1$ in which case, we get \begin{eqnarray*}
- \cW_\gb(v) + \gl \big| \{ v =-1\} \cap {\widehat \bbW}^2 \big| \leq
- \tau^* |v|^{(d-1)/d} + \gl \big| {\widehat \bbW}^2 \big|
\leq - \tau^* + \gl < 0 \, ,
\end{eqnarray*}
or $|v| < 1$ and \eqref{Wulff variational} applies \begin{eqnarray*}
- \cW_\gb(v) + \gl \big| \{ v =-1\} \cap {\widehat \bbW}^2 \big| \leq
- \tau^* |v| + \gl |v| < 0 \, .
\end{eqnarray*}
\end{proof}

\begin{rem}
As a consequence of the proof of Proposition \ref{infinite volume},
we see that the inverse of the spectral gap associated to the Glauber dynamics in the cube $[-N,N]^2$ grows faster than
$\frac{N}{(\log N)^\gk}$.
\end{rem}

\section{A one dimensional birth and death process for the droplet evolution}
\label{sec: toy model}

In this section we discuss a simple one dimensional toy model which mimics the random evolution of the volume of a droplet of the minus phase in a large cube of side $N$ in $\bbZ^d$ under a Glauber dynamics with plus boundary condition. The model goes as follows. Let $\a :=\frac{d-1}{d}$ and consider a birth and death process on the integers $\L :=\{0,\dots, N^d\}$, reversible with respect to the measure $$
\mu(x) := \frac{1}{Z} \exp(- x^{\a})
$$
and with birth rate $b(x) = (x\mmax 1)^\a, \; x<N$. By reversibility the death rate $d(x)$ is given by $$
d(x+1) := (x\mmax 1)^\a \exp((x+1)^\a -x^\a),\; x>0
$$
One easily checks that the drift given by $b(x)-d(x)$ is negative and proportional to $\a x^{2\a-1}$ for large $x$. The connection with the evolution of a large droplet of the minus phase under the Glauber dynamics with plus boundary condition in a large cube of side $N$ in $\bbZ^d$ is as follows. The variable $x$ represents the volume of the droplet at time $t$ which is assumed to form a compact set without holes. The rate $b(x)$ should then be interpreted as the rate with which a plus spin just outside the boundary of the droplet flips to minus one and gets attached to the droplet while the rate $d(x)$ represents just the opposite process in which a minus spin at the boundary flips to plus one and gets detached from the droplet. Clearly both these rates should be proportional to the size of the boundary which, for roundish shape, is of order of $x^\a$. Finally the drift comes from the reversibility condition together with the fact that the equilibrium distribution of the droplet volume should behave like the measure $\mu(x)$ above according to the results of section \ref{sec: Large Deviations}. Quite nicely the drift one gets out of these natural hypotheses is of the same order of that prescribed by an evolution by mean curvature $$
  \frac{d}{dt} x^{1/d} = - \frac{1}{x^{1/d}}
$$
Our goal now is to compute the precise asymptotic as $N\to \infty$ of the inverse spectral gap $S(N,d)$ and logarithmic Sobolev constant $L(N,d)$ of the above process in order to test the accuracy of the bounds proved in section \ref{subsec: Main Results}.

\begin{thm}
For any $d \ge2$ there exists a positive constant $k=k(d)$ such that
\begin{enumerate}[{\tt (i)}]
\item  \hskip 2cm $ \frac{N^2}{k} \leq L(N,d) \leq k\ N^2, \qquad \forall d\geq 2 $ \\ \item{} \hskip 2cm $ \frac{N}{k} \leq S(N,2) \leq kN $ \\
\item{} \hskip 3cm  $ S(N,d) \leq k,     \qquad  \forall d\geq 3 $
\end{enumerate}
\end{thm}

\proof We apply the method of Hardy inequalities envisaged in
\cite{Mi}) in order to compute sharp upper and lower bounds on the quantities of interest. We begin with the inverse spectral gap and define
\begin{align*}
      B_+(i) &:= \sup_{x>i} \Bigl(\,\sum_{y=i+1}^x
      \frac{1}{\mu(y)b(y)}\,\Bigr)\sum_{y\ge x}\mu(y) \\
      B_-(i) &:= \sup_{x<i} \Bigl(\,\sum_{y=x}^{i-1}
      \frac{1}{\mu(y)b(y)}\,\Bigr)\sum_{y\le x}\mu(y) \\
      B &:= \inf_{i\in \bbZ}\Bigl(\,B_+(i) \mmax B_-(i)\,\Bigr)
\end{align*}
The measure $\mu$ is extended on $\bbZ$ by setting $\mu(x) =0$ if $x \not \in \{ 0, \dots, N^d \}$. Then we have (see Proposition 1.3 of (\cite{Mi}) $$
     \frac{B}{2}  \leq S(N,d) \leq 4B
$$
Part ({\tt ii}) and ({\tt iii}) of the theorem follow at once from the simple estimates \begin{align}
 \sum_{y\ge x}\mu(y) &\approx x^{1-\a}\,\exp(-x^\a) \nonumber \\  \sum_{y=i+1}^x \frac{1}{\mu(y)b(y)} &\approx x^{1-2\a}\,\exp(x^\a)
\label{approx}
\end{align}
where $A\approx B$ means that there exists a universal constant $k$ such that $\frac{1}{k}\leq \frac{A}{B} \leq k$. We get in fact that for $i \in \{ 0, \dots, N^d \}$;
$B_+(i) \approx N$ for $d=2$ and $B_+(i) \leq k$
uniformly in $N$ for $d\geq 3$, while $B_-(i) \approx i^{1-2\a}\exp(i^\a)$ for any $d$. Notice that $B_+(i) = \infty$ if $i<0$ and $B_-(i) = \infty$ if $i> N^d$.

We now turn to the analysis of the logarithmic Sobolev constant. We define \begin{align*}
      A_+(i) &:= \sup_{x>i} \Bigl(\,\sum_{y=i+1}^x
      \frac{1}{\mu(y)b(y)}\,\Bigr)\log\Bigl(\frac{1}{\sum_{y\ge x}\mu(y)}\Bigr)
                                          \sum_{y\ge x}\mu(y) \\
      A_-(i) &:= \sup_{x<i} \Bigl(\,\sum_{y=x}^{i-1}
      \frac{1}{\mu(y)b(y)}\,\Bigr)\log\Bigl(\frac{1}{\sum_{y\le x}\mu(y)}\Bigr)
                                           \sum_{y\le x}\mu(y) \\
      A &:= \inf_{i\in \bbZ}\Bigl(\,A_+(i) \mmax A_-(i)\,\Bigr)
\end{align*}
Then we have (see Proposition 3.1 of \cite{Mi})
$$
     \frac{1}{20} A  \leq L(N,d) \leq 20 A
$$
and part ({\tt i}) follows at once from the bounds (\ref{approx}).

\end{document}